\documentclass[aps,prd,amsmath,amssymb,superscriptaddress,nofootinbib]{revtex4-2}

\usepackage{graphicx} 
\usepackage{dcolumn}  
\usepackage{bm}       
\usepackage{physics}  
\usepackage{hyperref} 
\usepackage{booktabs}
\begin{document}

\title{Solar Flares as a Probe of Neutrino Nature: Distinguishing Dirac and Majorana via Resonant Spin-Flavor Precession}

\author{David Delepine}
\email{delepine@ugto.mx}
\affiliation{Divisi\'on de Ciencias e Ingenier\'ias, Universidad de Guanajuato, C.P. 37150, Le\'on, Guanajuato, M\'exico.}

\author{A. Yebra}
\email{azarael@fisica.ugto.mx}
\affiliation{Divisi\'on de Ciencias e Ingenier\'ias, Universidad de Guanajuato, C.P. 37150, Le\'on, Guanajuato, M\'exico.}

\date{\today}

\begin{abstract}
 Resonant Spin-Flavor Precession (RSFP) of solar neutrinos is studied using the quantum density matrix formalism, explicitly taking into account collisional decoherence and solar matter density profiles. The transition probabilities for standard $^8$B solar neutrinos ($E \approx 10$ MeV) and ultra-high-energy flare neutrinos ($E \gtrsim 1$ GeV) under three magnetic field hypotheses: core-concentrated (Wood-Saxon), tachocline-confined (Gaussian), and turbulent convective (Power Law) are compared.  For standard LMA parameters, we show the resonance for 10 MeV neutrinos is strictly confined to the deep solar core ($r < 0.2 R_\odot$), rendering standard solar neutrinos insensitive to outer magnetic fields. Conversely, for 1 GeV flare neutrinos, the resonance shifts to the tachocline and convective zones, where strong fields ($B \sim 50$ kG) drive efficient spin conversion.
 We apply this effect to compute the difference between Dirac or Majorana neutrino scattering cross section as  electron-neutrino scattering and Coherent Elastic Neutrino-Nucleus Scattering (CE$\nu$NS). We show that significant asymmetry in these cross section are possible allowing in case of detection to distinguish between Dirac or Majorana neutrinos. In case of null observation, we show that this method can potentially improved the limit on the neutrino magnetic moment by one order to magnitude compared to current limits. 
\end{abstract}

\maketitle


\section{Introduction}

Since 1998 \cite{Super-Kamiokande:1998kpq, KamLAND:2002uet, SNO:2002tue}, the observation and confirmation of neutrino oscillations has firmly established that neutrinos are massive particles, yet their fundamental nature, Dirac or Majorana fermion\cite{Dirac:1928hu,Weyl:1929fm,Pauli:1930pc, Fermi:1934hr,Majorana:1937vz,Bilenky:1987ty},  remains one of the most significant open questions in particle physics. Usually, it is assume that the main smoking gun to determine the fundamental nature of the neutrino is the neutrinoless double beta decay ($0\nu\beta\beta$)\cite{Racah:1937qq, Furry:1939qr,  Pontecorvo:1957qd,Schechter:1981bd}. 
But the electromagnetic properties of the neutrino\cite{Fujikawa:1980yx, Schechter:1981hw, Pal:1981rm,Nieves:1981zt,Shrock:1982sc,Kayser:1982br,Vogel:1989iv,Giunti:2014ixa} allows to look for alternative method to distinguish between Dirac or Majorana fermions. As it has been showed previously in references \cite{Cisneros:1970nq, Okun:1986hi, Lim:1987tk, Akhmedov:1988uk,Minakata:1988ta, Pulido:1990bn,Balantekin:1990jb, Raffelt:1990yz, Akhmedov:2002mf, Ando:2003ie, Voloshin:2004vk},if neutrinos possess a non-zero magnetic moment $\mu_\nu$, their interaction with intense astrophysical magnetic fields can induce \textit{Resonant Spin-Flavor Precession} (RSFP). This mechanism rotates the spin of a left-handed electron neutrino ($\nu_{eL}$) into a right-handed state. The phenomenological consequence of this transition can reveal the nature of the neutrinos:
\begin{itemize}
    \item \textbf{Dirac Case:} The final state is a right-handed sterile neutrino ($\nu_{\mu R}$ or $\nu_{\tau R}$), which does not participate in standard weak interactions, leading to an observable disappearance of active flux (sterility).
    \item \textbf{Majorana Case:} The final state is an active right-handed antineutrino ($\bar{\nu}_{\mu}$ or $\bar{\nu}_{\tau}$), which continues to interact via the weak force, albeit with  cross-sections that can be slightly different.
\end{itemize}

Historically, the Sun has been used  as a natural laboratory for testing this mechanism \cite{Cisneros:1970nq, Lim:1987tk, Okun:1986hi, Raffelt:1999tx,Akhmedov:1999uz, Pulido:1992vb, Balantekin:2006sw}.However, standard solar neutrino studies (focusing on the MeV-scale $^8$B flux) have been strongly constraints by BOREXINO results\cite{Borexino:2008gab,Montanino:2008par,Borexino:2017fbd}. Experimental constraints from Borexino have effectively ruled out significant spin-flavor conversion in the standard energy range\cite{Borexino:2017fbd}. 

In this work, we propose a new  approach to distinguishing Dirac and Majorana neutrinos combining the observation of  High-Energy Solar Neutrinos ($E_\nu \gtrsim 1$ GeV) produced during solar flares and neutrino scattering processes on Earth. Unlike their MeV counterparts, GeV-scale neutrinos encounter the RSFP resonance condition in the Sun's outer layers—specifically the \textit{tachocline} and \textit{convective zone}. These regions are known to host strong magnetic fields ($B \sim 10^3 - 10^5$ G) generated by the solar dynamo models\cite{Spiegel:1992,Basu:1997zz,Charbonneau:2010zz}.

We use the quantum density matrix formalism to  model the neutrino evolution, explicitly accounting for collisional decoherence and magnetic turbulence in the solar plasma. We demonstrate that while standard solar neutrinos are insensitive to these outer fields, flare neutrinos undergo significant resonant precession. This leads to distinctive experimental signatures in neutrino scattering process like  or   Coherent Elastic Neutrino-Nucleus Scattering (CE$\nu$NS). These signatures provide a method for identifying the fundamental nature of the neutrino using future multi-messenger observations to detect solar flares neutrinos.

\section{Theoretical Framework}
\label{sec:theory}

To  describe the propagation of neutrinos through the turbulent solar environment, we treat the neutrino ensemble as an open quantum system\cite{ Burgess:1996mz, Balantekin:2004tk,Studenikin:2018vnp}. Unlike the standard Schrödinger wave-function approach, which assumes perfect coherence, the density matrix formalism allows us to explicitly account for decoherence effects induced by collisional scattering and magnetic field fluctuations.

\subsection{The Lindblad Master Equation}

The time evolution of the system is governed by the Lindblad master equation for the density matrix $\rho(t)$:
\begin{equation}
    \frac{d\rho}{dt} = -i [H_{\text{SFP}}, \rho] - \mathcal{D}[\rho],
    \label{eq:lindblad}
\end{equation}
where $H_{\text{SFP}}$ is the effective Hamiltonian driving the coherent evolution, and $\mathcal{D}[\rho]$ is the dissipator functional describing the irreversible loss of quantum coherence\cite{Lisi:2000zt, Friedland:2000rn,Oliveira:2019hyn,Kurashvili:2017zpz}.

The effective Hamiltonian for Resonant Spin-Flavor Precession (RSFP) in the flavor basis $(\nu_{eL}, \nu_{xR})^T$ is given by:
\begin{equation}
    H_{\text{SFP}} = \begin{pmatrix} 
    V_e - \delta c_{2\theta} & \mu_\nu B_\perp(r) \\ 
    \mu_\nu B_\perp(r) & V_x + \delta c_{2\theta} 
    \end{pmatrix},
    \label{eq:hamiltonian}
\end{equation}
where $\delta = \Delta m^2 / 4E$ is the vacuum oscillation term, and $c_{2\theta} = \cos 2\theta$. The matter potentials $V_e$ and $V_x$ are defined as:
\begin{align}
    V_e &= \sqrt{2}G_F (N_e - N_n/2), \\
    V_x &= 
    \begin{cases} 
     0 & \text{for Dirac (sterile } \nu_R\text{)}  \\
     0 & \text{for Majorana (active } \bar{\nu}_{\mu}\text{)} 
    \end{cases}
\end{align}
Note that for the Majorana case, we neglect small neutral current corrections to the antineutrino potential.

\subsection{Spin-Flavor Polarization Formalism}

 In our case,  the density matrix is written in terms of the polarization vector $\vec{P}(t)$ in the basis of Pauli matrices $\vec{\sigma}$:
\begin{equation}
    \rho(t) = \frac{1}{2} \left[ P_0(t) \mathbb{I} + \vec{P}(t) \cdot \vec{\sigma} \right].
    \label{eq:rho_expansion}
\end{equation}
Here, the vector $\vec{P} = (P_1, P_2, P_3)$ encodes all physical observables. Specifically, the longitudinal component $P_3$ corresponds to the helicity asymmetry of the ensemble related to the paralele component of the neutrino spin ($S_{\parallel}$):
\begin{equation}
    S_{\parallel}(r) \equiv P_3(r) = \frac{N_R - N_L}{N_R + N_L}.
\end{equation}
Substituting Eq.~(\ref{eq:rho_expansion}) into the Lindblad equation (Eq.~\ref{eq:lindblad}) yields the Generalized Bloch Equation for the polarization vector:
\begin{equation}
    \frac{d\vec{P}}{dt} = \vec{\Omega} \times \vec{P} - \Gamma \vec{P}_\perp.
    \label{eq:bloch_vector}
\end{equation}
This equation describes the precession of the neutrino spin around an effective magnetic field vector $\vec{\Omega}$, damped by a decoherence rate $\Gamma$. The components of the effective field are:
\begin{align}
    \Omega_1 &= 2\mu_\nu B_\perp(r) \quad \text{(Larmor torque)}, \\
    \Omega_2 &= 0, \\
    \Omega_3 &= V_{\text{eff}}(r) - \frac{\Delta m^2}{2E} \cos 2\theta \quad \text{(Matter resonance term)}.
\end{align}
The term $\Gamma \vec{P}_\perp$ phenomenologically accounts for the depolarization caused by magnetic turbulence in the convective zone, causing the vector magnitude to shrink ($|\vec{P}| < 1$) as the ensemble loses coherence.

\subsection{Adiabaticity and Resonance}

\begin{figure}
    \centering
    \includegraphics[width=0.9\linewidth]{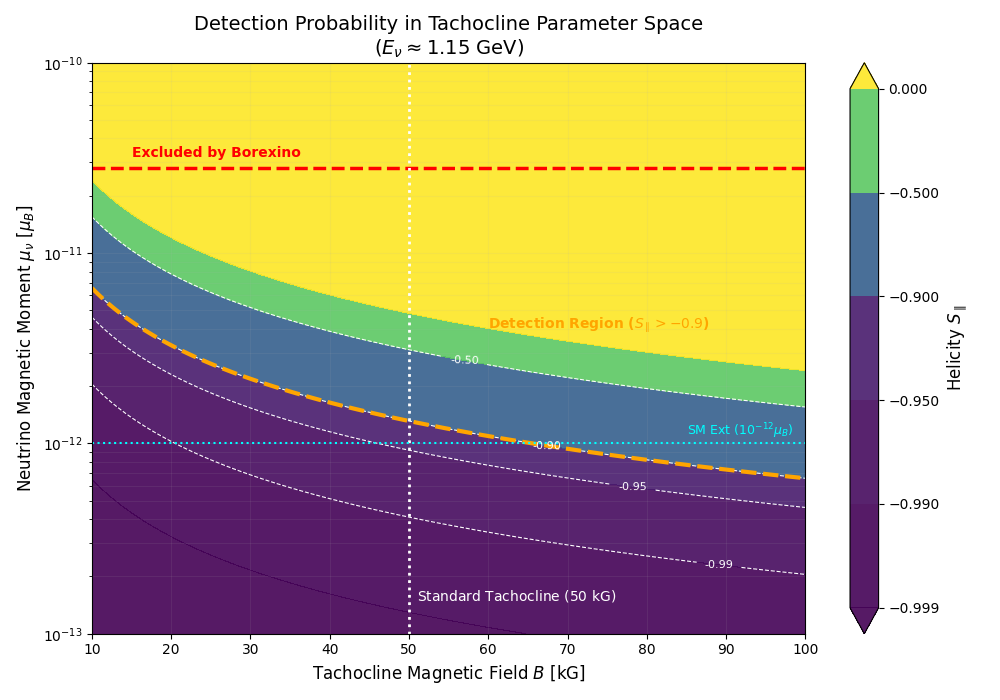}
    \caption{Contour plot of the neutrino helicity parameter $S_{\parallel}$ in the ($B$, $\mu_\nu$) parameter space, calculated for 1.15 GeV flare neutrinos resonating in the solar tachocline. The orange dashed line marks the threshold for a detectable signal, $S_{\parallel} > -0.9$. The red dashed line indicates the current upper limit on the neutrino magnetic moment from Borexino, $\mu_\nu < 2.8 \times 10^{-11} \mu_B$. The cyan dotted line represents a typical magnetic moment value predicted by some Standard Model extensions, $\mu_\nu = 10^{-12} \mu_B$. The vertical white dotted line marks the standard estimate for the peak tachocline magnetic field, $B \approx 50$ kG. }
    \label{munuB}
\end{figure}

The efficiency of the spin-flavor conversion is determined by the adiabaticity of the transition\cite{Balantekin:1990jb, Pulido:1990bn, Likhachev:1990ki, Guzzo:1991hi, Nunokawa:1996qr,Akhmedov:2003fu}. In the density matrix formalism, a transition is adiabatic if the precession frequency $|\vec{\Omega}|$ is much larger than the rate of change of the effective field direction. This is quantified by the adiabaticity parameter $\gamma$ at the resonance point ($\Omega_3 = 0$):
\begin{equation}
    \gamma = \frac{\Omega_1^2}{|\dot{\Omega}_3|} = \frac{(2\mu_\nu B_\perp)^2}{|\frac{d}{dr}(V_{\text{eff}})|_{\text{res}}}.
\end{equation}
The resulting non-adiabatic jump probability $P_{LZ}$ is given by the standard Landau-Zener formula\cite{Landau:1932,Zener:1932ws, Majorana:1932, Parke:1986jy, Petcov:1987cd, Kuo:1989qe}:
\begin{equation}
    P_{LZ} = \exp\left( -\frac{\pi}{2} \gamma \right)
\end{equation}
$P_{LZ}$ is related to $S_{\parallel}$ by:
\begin{equation}
S_{\parallel} = 1 - 2\exp\left(-\frac{\pi}{2}\gamma\right)
\end{equation}
\begin{itemize}
    \item \textbf{Adiabatic Limit ($\gamma \gg 1$):} The jump probability $P_{LZ} \to 0$, implying that the polarization vector perfectly follows the effective field $\vec{\Omega}$, leading to a complete reversal of helicity ($S_{\parallel}$ from $-1$ to $+1$).
    \item \textbf{Non-Adiabatic Limit ($\gamma \ll 1$):}  The jump probability $P_{LZ} \to 1$, and the helicity remains ``frozen'' at $S_{\parallel} \approx -1$,
\end{itemize}
The helicity state is governed by the adiabaticity parameter $\gamma$ at the resonance point. So if we want to have a difference of minimum 10\% in $S_{\parallel}$, assuming the neutrino energy to be at resonant point, it implies a limit on the product of neutrino magnetic fields and solar magnetic field, as given by

$$S_{\parallel} \approx 1 - 2 P_{flip} \implies -0.9 = 1 - 2 P_{flip} \implies P_{flip} = 0.95$$
Using the Landau-Zener approximation for the transition probability,setting $P_{flip} \ge 0.95$ yields a required adiabaticity of $\gamma \gtrsim 0.033$.Substituting the definition of $\gamma$:
$$\gamma = \frac{(2\mu_\nu B)^2}{|\frac{d}{dr}V_{eff}|_{res}} \gtrsim 0.033$$
Using the density gradient $|\frac{dV_{eff}}{dr}|$ specific to the solar tachocline model used in your simulations, we can quantify this product
$$\mu_\nu B \gtrsim 5.0 \times 10^{-8} \, \mu_B \cdot \text{G}$$
Assuming $B=50 kG$ as usual in Tachocline solar model. it means that
$$\mu_\nu  \gtrsim 1 \times 10^{-12} \, \mu_B
$$
From that, one can conclude that if no RSWP effects are observed with neutrino with Energy at resonant SFP point, the constraint to neutrino magnetic moment could be improve by more that one order of magnitude compared to present Borexino limit(see figure(\ref{munuB}).

\section{Application to Solar Neutrinos}
\label{sec:solar_application}

The efficiency of Resonant Spin-Flavor Precession (RSFP) depends critically on the interplay between the solar matter potential and the internal magnetic field profile and neutrino electromagnetic properties. In this section, we analyze the solar matter and solar magnetic field dependence for standard solar neutrinos ($^8$B flux) versus high-energy flare neutrinos.

\subsection{Solar Magnetic Field Models}

Since the internal magnetic structure of the Sun cannot be measured directly,  three distinct profiles widely used in the literature are used \cite{Solanki:2006zz, Hughes:2007book,Basu:2016zz, Fan:2009zz,Hathaway:2010zz,Charbonneau:2020zz}:

\subsubsection{Model I: Wood-Saxon Profile (Radiative Zone)}
This profile assumes a strong relic field frozen into the solar core ($r < 0.7 R_\odot$)\cite{Couvidat:2002,Friedland:2002ph,Miranda:2000bi}. It is parameterized by a Fermi-Dirac distribution:
\begin{equation}
    B(r) = \frac{B_0}{1 + \exp\left[ k \left( \frac{r}{R_\odot} - R_{\text{cut}} \right) \right]},
    \label{eq:wood_saxon}
\end{equation}
where $B_0$ is the peak field strength, $R_{\text{cut}} \approx 0.71 R_\odot$, and $k=20$. 

\subsubsection{Model II: Gaussian Tachocline Profile}
Standard solar dynamo theory suggests that strong toroidal fields are generated in the tachocline, the shear layer between the radiative and convective zones\cite{Spiegel:1992,Basu:1997zz,Charbonneau:2010zz}. We model this as a Gaussian peak centered at $r_{\text{tach}} \approx 0.71 R_\odot$:
\begin{equation}
    B(r) = B_{\text{max}} \exp\left[ - \frac{(r - r_{\text{tach}})^2}{2\sigma^2} \right].
    \label{eq:gaussian}
\end{equation}
With peak fields of $B_{\text{max}} \sim 20-50$~kG, this narrow region ($\sigma \approx 0.02 R_\odot$) can act as a potent ``spin flipper'' if the resonance condition is met locally.

\subsubsection{Model III: Convective Zone Power-Law}
In the outer convective zone ($r > 0.71 R_\odot$), the field is weaker and turbulent\cite{Kitchatinov:1993,Fan:2009zz}. Usually,  a power-law decay is assumed in this case:
\begin{equation}
    B(r) = B_s \left( \frac{R_\odot}{r} \right)^n, \quad \text{for } r > R_{\text{cut}}.
    \label{eq:power_law}
\end{equation}
To account for magnetohydrodynamic (MHD) turbulence, we decompose the field into a mean component $\langle B \rangle$ and a fluctuating component $\delta B$, which enters our formalism via the decoherence parameter $\Gamma$.

\subsection{Constraints from Borexino ($E \sim 10$ MeV)}

\begin{figure}
    \centering
    \includegraphics[width=0.9\linewidth]{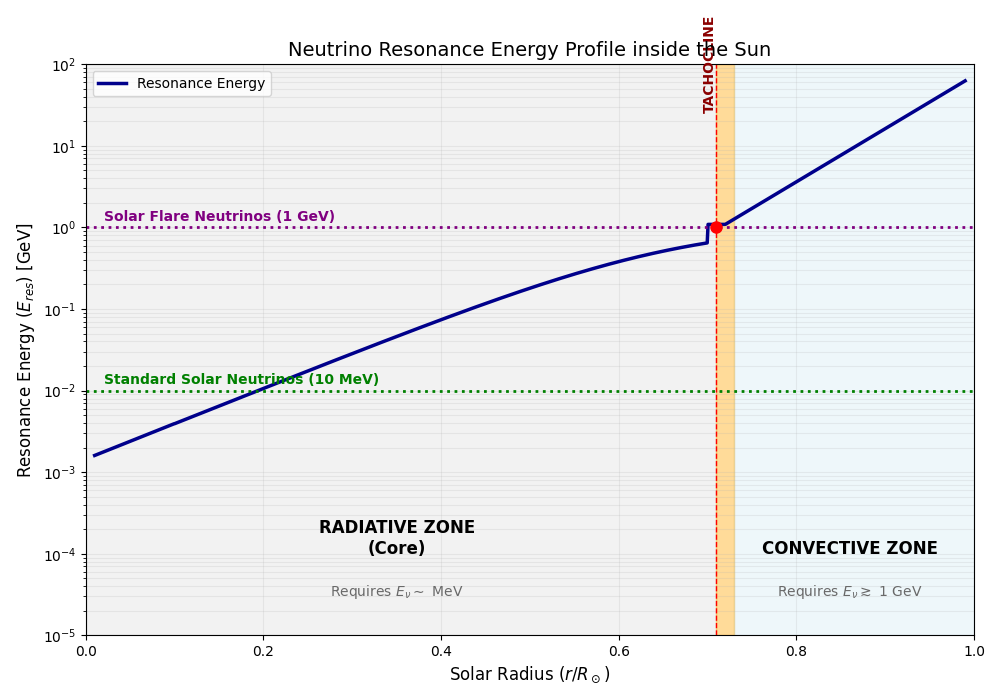}
    \caption{ Neutrino resonance energy ($E_{res}$) as a function of solar radius, calculated for standard Large Mixing Angle (LMA) oscillation parameters ($\Delta m^2 \approx 7.5 \times 10^{-5}$ eV$^2$). The plot illustrates the neutrino energy needed to fullfill the Resonant Spin-Flavor Precession (RSFP) condition across different solar regions. The horizontal green dotted line marks the energy of standard $^8$B solar neutrinos ($E \approx 10$ MeV), showing that their resonance is confined to the high-density radiative core ($r < 0.2 R_\odot$), where the adiabatic transition efficiency is suppressed. The horizontal purple dotted line indicates the energy of solar flare neutrinos ($E \sim 1$ GeV). The vertical red dashed line and orange shaded region highlight the tachocline ($r \approx 0.71 R_\odot$), the interface between the radiative and convective zones, where the resonance for 1 GeV neutrinos coincides with strong toroidal magnetic fields, facilitating efficient spin-flavor conversion.}
    \label{resonant}
\end{figure}

Current experimental data provides a strict boundary condition for our models.The non-observation of solar antineutrinos ($\bar{\nu}_e$) by Borexino limits the transition probability to $P(\nu_e \to \bar{\nu}) < 1.3 \times 10^{-4}$ (90\% C.L.)\cite{Borexino:2017fbd}. In terms of helicity, this implies:
\begin{equation}
    S_{\parallel} \lesssim -0.99974.
\end{equation}
This "frozen" state ($S_{\parallel} \approx -1$) indicates that for standard MeV-scale neutrinos, the resonance condition is either never met or the transition is highly non-adiabatic ($\gamma \ll 1$) in the regions where strong fields exist. Consequently, standard solar neutrinos effectively constrain the product $\mu_\nu B_{\text{core}} \lesssim 10^{-4} \mu_B \text{G}$, but they are "blind" to the strong fields in the outer tachocline and convective zones due to the density mismatch.
In figure(\ref{resonant}), we can see the relation between resonant neutrino energy as a function of distance from the center of the sun assuming the solar standard density distribution.

\subsection{The High-Energy Solution ($E \sim 1$ GeV)}

Neutrinos produced during solar flares ($E_\nu \sim 1$ GeV) offer a unique advantage:
\begin{enumerate}
    \item \textbf{Resonance Shift:} The resonance density scales as $\rho_{\text{res}} \propto \Delta m^2 / E_\nu$.For 1 GeV neutrinos, the resonance condition moves outward from the core into the \textit{tachocline} and \textit{convective zone}, precisely where the magnetic fields are strongest ($B \sim 10^3 - 10^5$ G).
   \item \textbf{Production Site:} Unlike core neutrinos, flare neutrinos are produced \textit{in situ} within the strong magnetic loops of the chromosphere, maximizing their exposure to the B-field immediately upon creation.
\end{enumerate}

\begin{figure}
    \centering
    \includegraphics[width=0.9\linewidth]{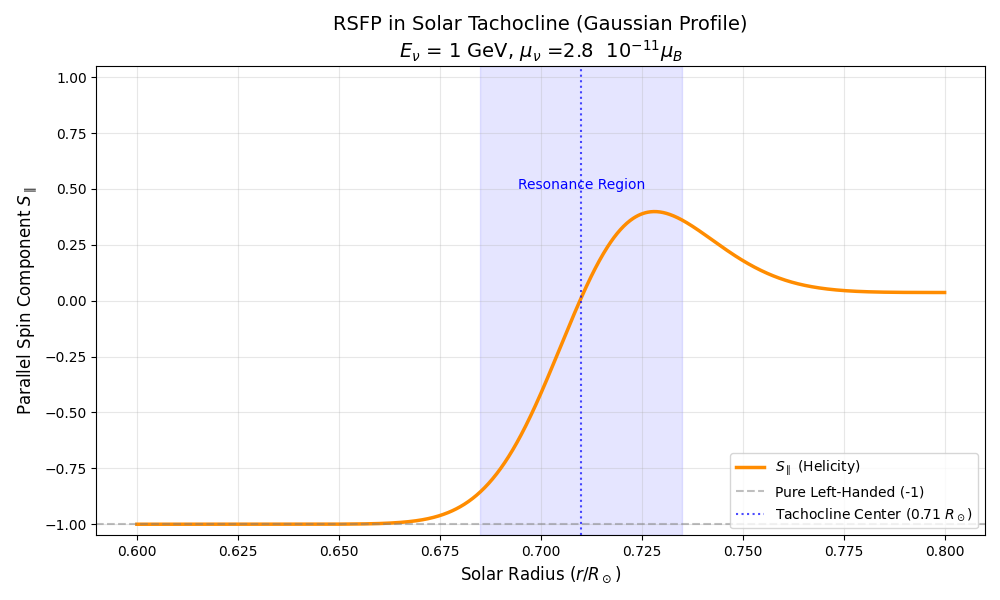}
    \caption{Evolution of the parallel spin component $S_\parallel$ for a 1 GeV solar flare neutrino traversing the solar tachocline. The simulation assumes a Gaussian magnetic field profile centered at $r = 0.71 R_\odot$ with a peak strength of $B = 50$ kG and a neutrino magnetic moment at the Borexino limit ($\mu_\nu = 2.8 \times 10^{-11} \mu_B$). The shaded blue area indicates the resonance region where the adiabaticity condition allows for a significant conversion from the initial left-handed state ($S_\parallel = -1$) to a mixed state, demonstrating the high efficiency of RSFP at this energy scale}
    \label{tach}
\end{figure}

\begin{figure}
    \centering
    \includegraphics[width=0.9\linewidth]{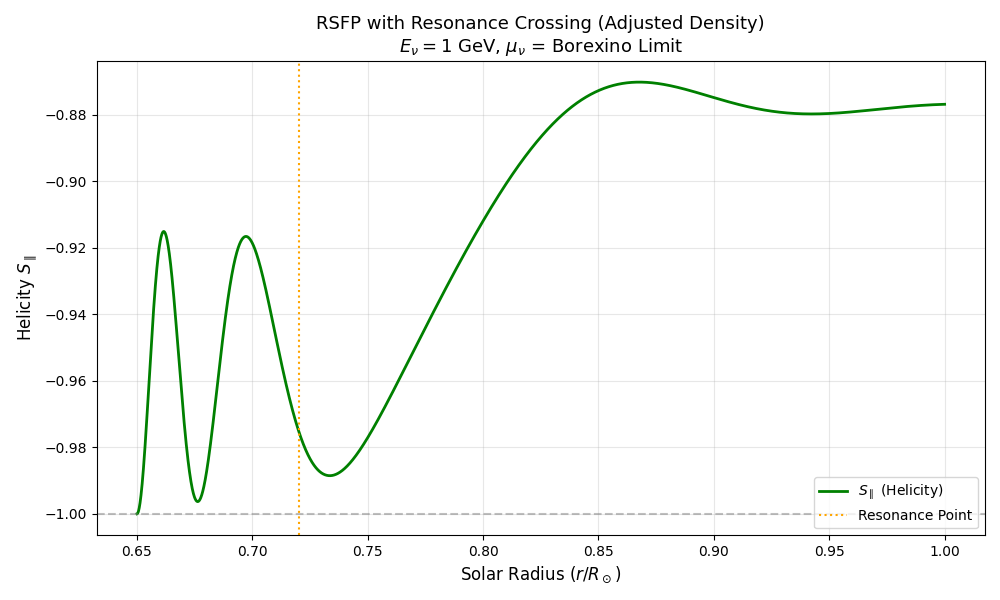}
    \caption{Helicity evolution ($S_\parallel$) for a 1 GeV neutrino propagating through the solar convective zone, using an adjusted density profile to explicitly capture the resonance crossing at $r \approx 0.72 R_\odot$ (marked by the vertical orange dotted line). The green curve shows the macroscopic deviation from the pure left-handed state ($S_\parallel = -1$) initiated by the resonance. The final stabilized value of $S_\parallel \approx -0.88$ illustrates that measurable depolarization effects are theoretically expected in the outer solar layers for high-energy neutrinos.}
    \label{convective}
\end{figure}

While MeV neutrinos remain unaffected, 1 GeV neutrinos in the Tachocline scenario can undergo a near-complete adiabatic conversion ($S_{\parallel} \to 0.04$) assuming the neutrino magnetic moment to be close to the Borexino limit as it can be seen in figure(\ref{tach},\ref{convective}). 

\section{Scattering Cross Sections and Experimental Signatures}
\label{sec:scattering}

In this section, we analyze the differential cross-sections for neutrino-electron scattering ($e-\nu$) and Coherent Elastic Neutrino-Nucleus Scattering (CE$\nu$NS). Our goal is to identify observable signatures that distinguish Majorana from Dirac neutrinos, particularly when Resonant Spin-Flavor Precession (RSFP) modifies the helicity state of high-energy solar flare neutrinos.

\subsection{Neutrino-Electron Scattering ($e-\nu$)}

For ultra-high-energy solar neutrinos ($E_{\nu} \gg m_{\nu}$), we work in the massless limit where chirality and helicity are effectively degenerate. The differential cross-sections for the scattering of a neutrino with an arbitrary parallel spin component $s_{\parallel}$ on an electron are given by\cite{Rosen:1982pj,Kayser:1982br, Bernabeu:2000hf}:

\begin{itemize}
    \item \textbf{Dirac Cross-Section ($\sigma_D$):}
    \begin{equation}
        \frac{d\sigma_D}{d\Omega}_{\nu e} \propto (1 - s_{\parallel}) \left[ Y(\theta) + Z(\theta) \right]
    \end{equation}
    
    \item \textbf{Majorana Cross-Section ($\sigma_M$):}
    \begin{equation}
        \frac{d\sigma_M}{d\Omega}_{\nu e} \propto 2 \left[ Y(\theta) - s_{\parallel} Z(\theta) \right]
    \end{equation}
\end{itemize}

Here, the kinematic functions $Y$ and $Z$ encapsulate the Standard Model weak couplings:
\begin{align}
    Y(\theta) &= (g_V^2 + g_A^2) \left[ 1 + \frac{(E_e + E_\nu \cos \theta)^2}{s} \right], \\
    Z(\theta) &= 2 g_V g_A \left[ 1 - \frac{(E_e + E_\nu \cos \theta)^2}{s} \right].
\end{align}
The vector and axial-vector couplings are $g_V = -\frac{1}{2} + 2\sin^2 \theta_W$ and $g_A = -\frac{1}{2}$.

It is crucial to note that for a pure left-handed neutrino ($s_{\parallel} = -1$), the Dirac prefactor $(1 - s_{\parallel})$ becomes 2. Similarly, the Majorana term becomes $2[Y + Z]$. Thus, in the absence of spin precession, $\sigma_D = \sigma_M$, and the two cases are experimentally indistinguishable. The distinction arises \textit{only} when RSFP induces a deviation from pure left-handedness ($s_{\parallel} > -1$).

We quantify this distinction using the Normalized Asymmetry $\mathcal{A}$:
\begin{equation}
    \mathcal{A} = \frac{\sigma_M - \sigma_D}{\sigma_M + \sigma_D}.
\end{equation}
For standard solar neutrinos, where Borexino limits imply $s_{\parallel} \approx -0.9997$, this asymmetry is negligible ($\mathcal{A} \approx 5.6 \times 10^{-5}$).


For GeV-scale neutrinos produced in solar flares, the resonance condition shifts to the outer solar layers, leading to macroscopic deviations in helicity.

\subsubsection{Scenario A: The Convective Zone (assuming $S_{\parallel} \approx -0.94$)}
Turbulence in the convective zone induces a mild depolarization. This results in an asymmetry of $\mathcal{A} \approx 1.1\%$. While an order of magnitude larger than the standard solar case, this remains challenging to resolve given current systematic uncertainties.

\subsubsection{Scenario B: The Tachocline (assuming efficient RSFP as $S_{\parallel} \approx -0.1$)}
This scenario represents a near-complete resonant transition at $r \approx 0.71 R_\odot$. The physical consequences are profound:
\begin{itemize}
    \item \textbf{Dirac Case:} The flux is converted into sterile right-handed neutrinos ($\nu_R$), which do not interact. The cross-section is suppressed by a factor of $\approx (1 - (-0.1))/2 = 0.55$.
    \item \textbf{Majorana Case:} The flux is converted into active right-handed antineutrinos ($\bar{\nu}$), which continue to scatter via the Weak interaction.
\end{itemize}
 This difference leads to a normalized asymmetry of 16.85\% (see Table~\ref{tab:sfp_results}), a "smoking gun" signal accessible to future high-precision detectors.

\begin{table}[ht]
\centering
\caption{Normalized cross-section difference ($\mathcal{A}$) between Majorana and Dirac neutrinos at $E_\nu = 1$ GeV for different solar magnetic field scenarios assuming neutrino magnetic moment given by Borexino limit and $S_{\parallel}$ as given in the text .}
\label{tab:sfp_results}
\begin{tabular}{@{}lccc@{}}
\toprule
\textbf{Scenario} & \textbf{Helicity ($S_{\parallel}$)} & \textbf{Asymmetry ($\mathcal{A}$)}  \\ \midrule
Convective Zone   & $-0.94$                        & $1.12\%$                            \\
Tachocline        & $0.04$                        & $16.85\%$                         \\ \bottomrule
\end{tabular}
\end{table}

\subsection{Coherent Elastic Neutrino-Nucleus Scattering (CE$\nu$NS)}

The CE$\nu$NS process offers an even more sensitive discriminator because it is flavor-blind and coherent. The differential cross-section with respect to nuclear recoil energy $T$ is given by\cite{Freedman:1973yd,Kopeliovich:1974qv,Drukier:1983gj, Horowitz:2003cz,Lindner:2016wqj, Kosmas:2017zlb, Bednyakov:2018mjd,Akimov:2017ade,Plestid:2020ods}:

\begin{itemize}
    \item \textbf{Dirac Neutrinos ($\sigma_D$):}
    \begin{equation}
        \frac{d\sigma_D}{dT} \propto (1 - s_{\parallel}) Q_w^2 F^2(T)
    \end{equation}
    
    \item \textbf{Majorana Neutrinos ($\sigma_M$):}
    \begin{equation}
        \frac{d\sigma_M}{dT} \propto 2 Q_w^2 F^2(T)
    \end{equation}
\end{itemize}
Here, $Q_w = N - (1 - 4\sin^2 \theta_W)Z$ is the weak nuclear charge,  $N$ is the number of neutrons and $Z$ is the number of proton and $F(T)$ is the nuclear form factor.

In the Majorana case, a spin-flip produces an active antineutrino ($\bar{\nu}$). Since the weak charge $Q_w$ is invariant under $CP$ transformation (ignoring radiative corrections), $\nu$ and $\bar{\nu}$ scatter with identical rates in CE$\nu$NS. Therefore, the Majorana cross-section is independent of helicity.

Conversely, in the Dirac case, the right-handed state is sterile. This leads to a direct loss of signal.
\begin{itemize}
    \item \textbf{Tachocline Prediction ($s_{\parallel} = 0.04$):} The $\mathcal{A}$ Dirac-Majorana asymmetry is around $-36\%$ and Dirac cross section is reduced by a factor of $ \approx 0.45$ compared to Majorana case.
    \item \textbf{Convective Prediction ($s_{\parallel} = -0.94$):} Even the mild depolarization of the convective zone yields a 3\% difference, significantly easier to resolve than the 1.1\% asymmetry in electron scattering.
\end{itemize}

\section{Conclusion}
\label{sec:conclusion}

In this work, we have established that combining the neutrino electromagnetic properties with solar magnetic field through RSFP process, can lead to different signatures between Dirac or Majorana neutrino measuring the scattering cross section at Earth. 

Our analysis reveals a fundamental dichotomy dictated by the neutrino energy scale:
\begin{enumerate}
    \item \textbf{The Standard Regime ($E \sim 10$ MeV):} For standard $^8$B solar neutrinos, the resonance condition is confined to the deep solar core. The null results from Borexino impose strong limit on the neutrino magnetic moment and confirms that standard solar neutrinos cannot currently distinguish neutrino nature via RSFP.
    
    \item \textbf{The High-Energy Regime ($E \sim 1$ GeV):} For neutrinos produced in solar flares, the resonance naturally shifts to the \textit{tachocline} and \textit{convective zone}. In these outer layers, the strong magnetic fields ($B \sim 10^3-10^5$ G) facilitate a near-adiabatic transition which means a relatively efficient RSFP. 
\end{enumerate}

We predict that for 1 GeV neutrinos resonating in the solar tachocline, the Resonant Spin-Flavor Precession (RSFP) induces a macroscopic deviation in the scattering cross-sections:
\begin{itemize}
    \item In \textbf{Neutrino-Electron Scattering}, we predict a normalized asymmetry of $\mathcal{A} \approx 37\%$ between the Dirac and Majorana scenarios.
    \item In \textbf{Coherent Elastic Neutrino-Nucleus Scattering (CE$\nu$NS)},  the Dirac model predicts a massive \textbf{45\% signal reduction} (due to conversion into sterile states), while the Majorana model predicts a stable flux (due to conversion into active antineutrinos).
\end{itemize}

The primary challenge in observing this effect is distinguishing transient solar flare neutrinos from the continuous isotropic atmospheric background. We propose that a multi-messenger strategy is essential for success. By utilizing real-time gamma-ray triggers from the High Altitude Water Cherenkov (HAWC) observatory, future neutrino detectors can define precise temporal windows for analysis, suppressing the atmospheric background by orders of magnitude.

By combining the high-statistics capabilities of future detectors like Hyper-Kamiokande\cite{Abe:2018uyc,Abe:2021cMw} and IceCube-Gen2\cite{Aartsen:2015yva,Aartsen:2020fgd,Abbasi:2021jcq} combined with $\gamma $ ray observatory as HAWC\cite{Abeysekara:2013tza,Abeysekara:2018qxs,Linden:2020bkz} with the flavor-blind sensitivity of CE$\nu$NS \cite{Akimov:2017ade,Akimov:2020pdx,Akimov:2021dab,Bonet:2020ntx,Hakenmuller:2019ecb,Aguilar-Arevalo:2019jlr,Agnolet:2016zir, Angloher:2019flc}, there is a open window to be able to distinguish between Dirac and Majorana neutrino. 

Furthermore, if no such effects are observed, our results imply that the constraint on the neutrino magnetic moment could be improved by more than one order of magnitude compared to the present Borexino limit.

\begin{acknowledgments}
We acknowledge financial support from SECIHTI and SNII (M\'exico).
\end{acknowledgments}

%

\end{document}